\documentclass[%
reprint,
superscriptaddress,
amsmath,amssymb,
prl,
]{revtex4-2}

\usepackage{ulem}%
\usepackage{graphicx}
\usepackage{dcolumn}
\usepackage{bm}

\newcommand{\kBT}{k_\mathrm{B}T}
\newcommand{\cc}{c_\mathrm{c}}

\usepackage{lineno}

\usepackage{color}
\renewcommand{\emph}[1]{\textit{#1}}

\begin{document}
	
	
	\title{Collective dynamics and elasto-chemical cluster waves in communicating colloids with explicit size response}

	
	\author{Nils G\"oth}
	\affiliation{Applied Theoretical Physics--Computational Physics, Physikalisches Institut, Albert-Ludwigs-Universit\"at Freiburg, D-79104 Freiburg, Germany}
	\author{Joachim Dzubiella}
	\email[Corresponding author: ]{joachim.dzubiella@physik.uni-freiburg.de}
	\affiliation{Applied Theoretical Physics--Computational Physics, Physikalisches Institut, Albert-Ludwigs-Universit\"at Freiburg, D-79104 Freiburg, Germany}
	\affiliation{Cluster of Excellence livMatS @ FIT--Freiburg Center for Interactive Materials and Bioinspired Technologies, Albert-Ludwigs-Universit\"at Freiburg, D-79110 Freiburg, Germany}
	
	\date{August 21, 2024}
	
\begin{abstract}
\textbf{Chemical communication, response, and feedback are key requirements for the function of adaptive materials with life-like properties. However, how communication on the single cell-level impacts the collective structural, dynamical and mechanical behavior of active soft matter is not well understood. Here, we report  how communication controls the spatiotemporal structure and phase behavior of active, hydrogel-based colloidal liquids using Brownian particle-based simulations with explicit resolution of the chemical signaling waves as well as the individual particle's elastic response to communication and crowding.  We find a rich topology of nonequilibrium active phases, vastly tuneable by the signaling magnitude, in particular, active melting, synchronization transitions from uncorrelated to antiphase oscillatory liquids, or to in-phase oscillations with accompanying elasto-chemical cluster waves. Our work employs minimal physical principles required for communication-mediated dynamics of microscopic, fluctuating systems, thus uncovering universal aspects in signaling soft systems.}
\end{abstract}
	
\maketitle
\small
	


\section{Introduction}
Modern materials science and engineering aims at the design of sustainable and adaptive living materials~{\cite{Niederholtmeyer2018, Rampioni2018, LiebchenBook, Rifaie-Graham2023, Bauerle2018,Nguyen2018, Danino2010}}, in particular, with a programmable or homeostatic (self-regulating) response~\cite{C6CS00738D}. Synthetically, this can be realized with catalytically active hydrogels which are soft, stimuli-responsive, and can be chemically fueled and excited~\cite{Wessling2021, Dhanarajan2002, Heckel, Blanc2024}. Mechanically actuated motion and entrainment can be generated by exploiting hydrogel (de)swelling in response to a periodic stimulus~\cite{Geher-Herczegh2024}. More complex self-regulatory behavior is obtained by adding chemo-mechano-chemical feedback mechanisms~\cite{Fusi, Yashin2012}. For achieving controllable communication, chemical feedback reaction networks are  compartmentalized into hydrogel spheres where fuel-driven pH fronts can be sent, received, and regulated~\cite{Maity2021}. 

Communication and collective self-organization are central to living systems. Biological cells or bacteria secrete and sense molecules to `talk' with each other to regulate their collective function and spatiotemporal structuring, for example, in biofilm formation or in cell migration~\cite{Miller2001, Lopez2009,QS, Bauerle2018}. In biological tissues, the soft cells use communication to control their active elastic response, such as mechanical deformations and oscillations. Cells can, for example, collectively increase and decrease their sizes in a sustained and locally synchronized fashion~\cite{Thiagarajan2022, Zehnder2015, Karma2013}. The interplay between synchronization, repulsion and deformation then leads to {(mechano-)chemical propagation of morphogenetic waves in tissues~\cite{Zhang2023, Heisenberg, Danino2010}} as in fly endoderms~\cite{morphowave,morphowave2}.

Existing approaches to model chemical communication in liquids are typically based on chemotactically or chemophoretically interacting self-propelled particles~\cite{Agudo-Canalejo2019, Ziepke2022,Fadda2023, Dinelli2023, Ridgway2023,PhysRevE.89.062316, Grauer2024,Illien2024, Mognetti2013}, leading to moving clusters~\cite{Agudo-Canalejo2019}, collective clustering~\cite{Ziepke2022, Fadda2023}, dynamical phase separation~\cite{Dinelli2023}, or oscillatory patterns~\cite{PhysRevE.89.062316,Ridgway2023}.  It was shown by a minimal model of chemical communication fields how particles collectively optimize static spatial order~\cite{Zampetaki2021}.  Active models for mechanical deformation waves use dense active matter systems with an intrinsic (pre-defined) pulsation of size~\cite{Tjhung2017,Zhang2023,Pineros2024}. However, the impact of chemical communication on the structure and its coupling to the mechanical response of soft liquids is not well understood, in particular, the physical connections of cellular level fluctuations and the collective dynamics are unclear.

Here, we report on the consequences of chemical communication on the structure and phases of dispersions of models of hydrogel-based {(non-motile)} responsive colloids~\cite{Paloli2013,Baul2021} using particle-based Brownian dynamics extended to include communication, response, and feedback. We explore in detail the physical principles behind the coupling of single-colloid mechanics, and communication with respect to the collective dynamics. In contrast to recent theoretical work on active pulsating matter~\cite{Zhang2023}, the cells, their mechanical response, and accompanying chemical fields are explicitly resolved and dynamically tracked in our simulations. Our approach is inspired by recent experimental developments in the field of synthetic, synergistic (not slaved) self-oscillating hydrogel colloids. These colloids are catalytically active and able to convert a constant energy supply into chemo-mechanical oscillations based on pH-response with negative feedback~\cite{Narita2013, Wessling2021}. We recently calculated the dynamical state diagram of such a colloid, with resemblance to a neuron cell~\cite{Milster2023}. Here, we demonstrate that oscillations, their synchronization, and dynamical patterns emerge as a result of collective interactions, precisely, through the elastic and chemical coupling among the units.

\begin{figure}[t]
	\centering
	\includegraphics[width=7.0cm]{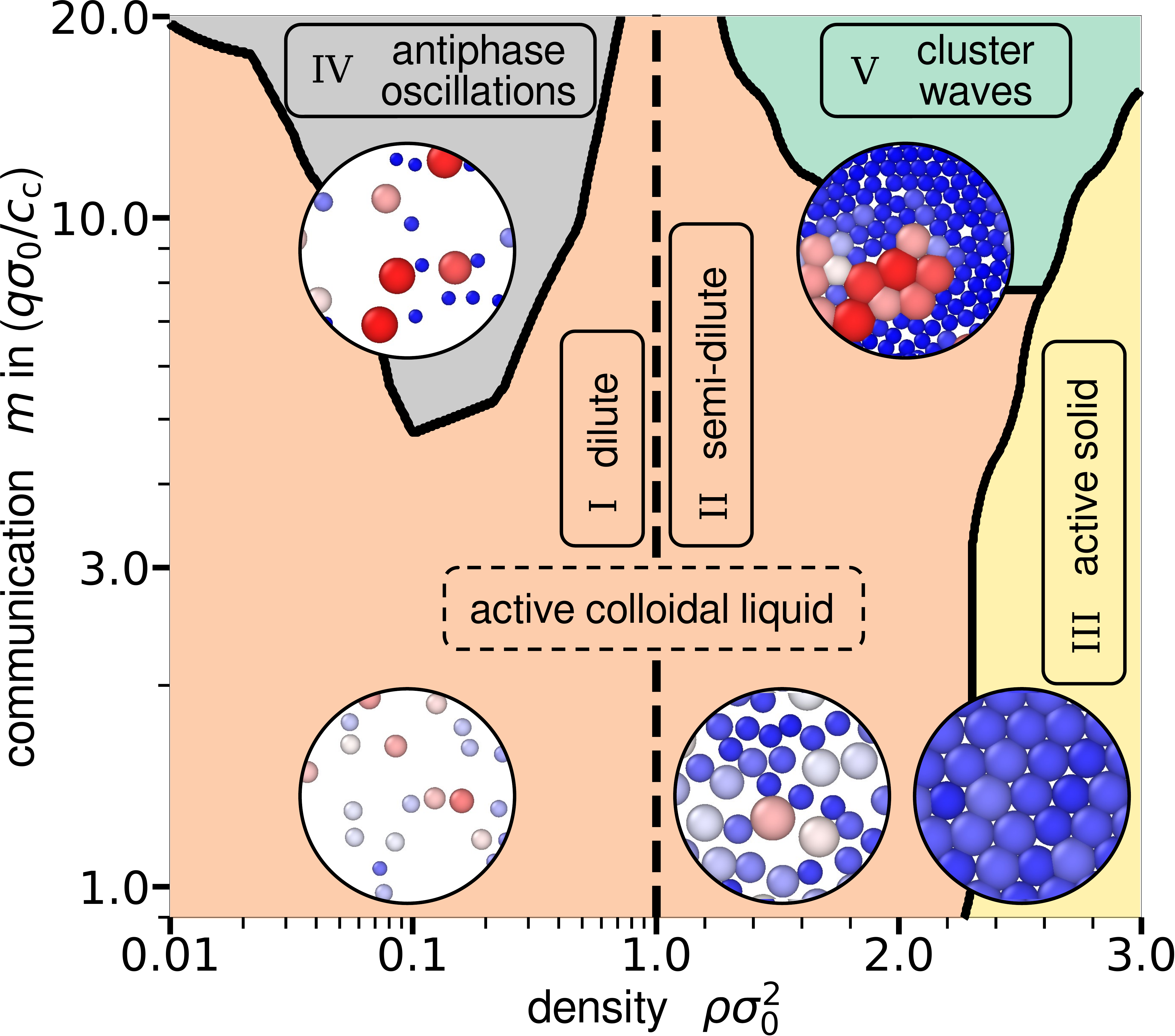}
	\caption{{\bf Dynamical state diagram in the communication-density space.} The collective dynamic behavior is governed by the (chemical) communication strength $m$, and by the mechanical/elastic interactions which are tuned by the (number) density $\rho$. We distinguish between five states: (I) active colloidal liquid in the dilute and (II) semi-dilute regime, (III) active colloidal solid, (IV) antiphase oscillations, and (V) cluster waves. The $\rho$-axis is logarithmic until $\rho\sigma_0^2=1.0$ and linear above for clearer presentation of the topology of the states. Related movies can be found in the Supplementary Movies S1-S5. All snapshots and movies are made with OVITO~\cite{OVITO}.}
	\label{fig.state_diagram}
\end{figure}

We report a rich nonequilibrium state diagram, shown in Fig.~\ref{fig.state_diagram}. In particular, we find not only communication-controlled active versions of the conventional dilute liquid (I), semi-dilute liquid (II), and solid (III) colloidal phase~\cite{Baul2021,Tsiok2022, Kuk2023}, but also two novel synchronized oscillatory phases, namely antiphase oscillations (IV) and in-phase oscillatory cluster waves (V) for denser and more correlated systems. The importance of density for oscillatory dynamics is experimentally demonstrated~{~\cite{Taylor2009,Taylor2015}}, for example, in yeast cells, which only show chemical (glycolytic) oscillations above certain cell density thresholds~\cite{DeMonte2007} tuneable by the signaling agents~\cite{Bier2000}. {Similarly, recent experiments with Belousov-Zhabotinsky reaction-slaved hydrogel beads present no oscillations for an isolated bead, while an assembly of beads oscillates chemically and mechanically~\cite{Blanc2024}}.

Overall, our work demonstrates how the behavior of conventional colloidal liquids changes if it is perturbed by active chemical communication. We show how systematically increasing the signaling magnitude does not only change the structure (like the radial distribution function) and the phase boundaries (such as liquid to solid) but also leads to new dynamic phases with tuneable oscillations and synchronizations. Coming from the colloidal physics route, we therefore uncover the basic physical principles and minimal ingredients for mechano-chemical coupling between chemical signaling and liquid viscoelasticity in active `living' liquids, featuring oscillations and cluster waves like in biological morphogenesis~\cite{Heisenberg, morphowave,morphowave2,Zhang2023}. Our work will also be inspiring for fundamental developments in nonequilibrium theoretical physics (of active matter), and serve for the rational design of synthetic living matter.



\begin{figure*}[bt]
	\centering
	\includegraphics[width=\textwidth]{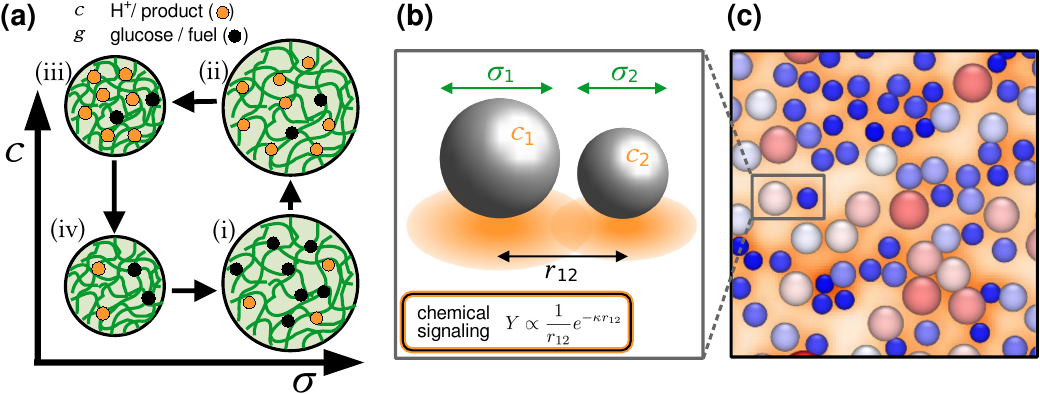}
	\caption{{\bf Model of excitable and chemically communicating colloids.} \textbf{(a)} Sketch of the oscillation cycle of one responsive and catalytically active colloid (i)-(iv): (i) Glucose (chemical fuel) enters the penetrable colloidal cell and at concentration $g$ leads to production of protons with inside concentration $c$; (ii) the responsive colloid responds to the protons and shrinks (towards lower size $\sigma$); (iii) the proton concentration $c$ decreases because of the low permeability of the collapsed dense (hydrogel) colloid to glucose; (iv) the low $c$ leads to a re-swelling of the cell. \textbf{(b)} Schematic representation of two active colloids in center-to-center distance $r_{12}$. Each particle has its individual diameter $\sigma_i$ and internal proton concentration $c_i$. The orange areas illustrate the chemical concentration fields (`clouds') of each particle with which they communicate via a Yukawa field $Y$. \textbf{(c)} Visualization of a small part of the 2D simulation box. The particles' colors indicate their size (large $\rightarrow$ red, small $\rightarrow$ blue), the orange background visualizes the total chemical field.}
	\label{fig.methods}
\end{figure*}

\section{Model}
In our work, a single active colloid is assumed to be governed  by the chemo-mechanical feedback cycle depicted in Fig.~\ref{fig.methods}(a). Such a reaction cycle can lead to stable synergistic self-oscillations in a narrow parameter range for mechanically bistable particles~\cite{Wessling2021, Milster2023}. In this work, we employ monostable colloids which do not display oscillations in the isolated case in order to examine the important question how particle-particle communication triggers oscillations, synchronization, and new collective dynamics.

We study dispersions of chemically communicating colloids in a quasi-2D monolayer setup. Each particle $i$ in our active colloidal dispersion has four degrees of freedom, cf.~Fig.~\ref{fig.methods}(b): two for the spatial coordinates $\boldsymbol{x}_i$, one for its diameter $\sigma_i$, and one for its inside proton concentration $c_i$. Their dynamics are coupled through (elastic) collisions and chemical communication, defined by:
\begin{subequations}
	\begin{eqnarray}
		\xi_x \dot{\boldsymbol{x}}_i &=& - \sum_{j \neq i} \nabla \Phi_{ij} + \sqrt{2\xi_x/(\beta \Delta t)} \boldsymbol{\eta}_{ix},
		\label{eq.model_eom_x}\\ \nonumber
		\xi_\sigma \dot{\sigma}_i &=& - \sum_{j \neq i} \partial_{\sigma} \Phi_{ij} - \partial_{\sigma} U \nonumber \\
		&& - m\left[ c_i + \sum_{j \neq i} Y_{ji} - c_\mathrm{c}\right] + \sqrt{2\xi_\sigma/(\beta \Delta t)} \eta_{i\sigma},
		\label{eq.model_eom_s}\\ 
		\dot{c}_i &=& k_0 c_\mathrm{c} e^{-B/\sigma_i^3} - k_l c_i.
		\label{eq.model_eom_c}
	\end{eqnarray}
\end{subequations}
The translational and particle size motions are modeled by Brownian dynamics, cf. Eq.~(1a) and (1b). Elastic interactions between the spherical cells are included through a purely repulsive Hertzian pair potential $\Phi_{ij}(r_{ij};\sigma_i,\sigma_j)$ as a function of pair distances $r_{ij}$ and instantaneous sizes $\sigma_i$ and $\sigma_j$. The latter fluctuate in a single-particle U-shaped FENE energy landscape, $U(\sigma_i)$, which is in first order harmonic, centered at the mean isolated-particle size, $\sigma_0$. This potential is not bistable, hence a single particle cannot oscillate~\cite{Milster2023}. The friction parameters $\xi_x$ and $\xi_\sigma$ (and corresponding noises $\boldsymbol{\eta}_{ix}$ and $\eta_{i\sigma}$, respectively) determine the intrinsic time scales of translation and size fluctuations at (inverse) temperature $\beta = 1/(\kBT)$. A simulation snapshot for many colloids is shown in Fig.~\ref{fig.methods}(c). See the Methods section for more details on potentials and simulations. 

The square bracket term in Eq.~(\ref{eq.model_eom_s}) introduces the effects of chemical signaling and communication to the $\sigma$-response. The communication prefactor $m$ defines the (chemical) coupling strength between the rate of size change and the total cellular proton concentration. The latter is a sum of the own concentration $c_i$ (self-feedback) and the one coming from all other particles, modeled by a Yukawa concentration field for colloid $j$, measured by colloid $i$, $Y_{ji} (r_{ij}; c_j, \sigma_j) = c_j \sigma_j e^{\kappa \sigma_j/2} e^{-\kappa r_{ij}} / (2r_{ij})$. The Yukawa cloud with communication length $\kappa^{-1}$ is the solution of the stationary diffusion equation with a source proportional to $c_j$ and a generic loss term (see Methods). We fix the communication length to $\kappa^{-1}= \sigma_0$ throughout the work. Overlapping Yukawa clouds lead to chemo-mechanical communication: A high total proton concentration (exceeding the reference concentration $c_c$) will cause the colloid to shrink. Conversely, when the proton concentration is low, the colloid will swell. The feedback cycle is closed by the proton production rate for colloid $i$, Eq.~(\ref{eq.model_eom_c}), including the mechanical permeation feedback of pH-responsive hydrogels~\cite{Wessling2021, Milster2023}.

\section{Results}
\subsection{Two-particle communication triggers stable antiphase oscillations}

\begin{figure*}[ht!]
	\centering
	\includegraphics[width=\textwidth]{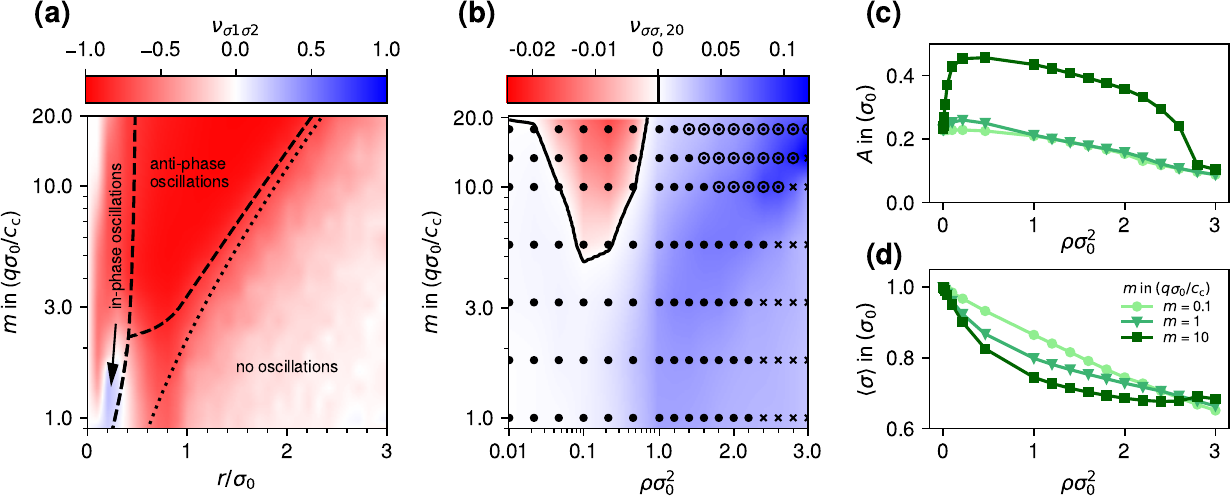}
	\caption{{\bf Size cross-correlations and order parameters in two- and many-body communication.} \textbf{(a)} Size-cross-correlations (Pearson coefficient $\nu_{\sigma_1\sigma_2}$) in the two-particle system in a log-lin heat map as a function of the $m$-value and the fixed interparticle distance $r$. The dashed line is from a numerical stability analysis of the system without noise (see Supplementary Sec.~II). The dotted line is from an analytical stability analysis (see Supplementary Sec.~III). \textbf{(b)} Local size-cross-correlations (local Pearson coefficient $\nu_{\sigma\sigma,20}$) as heat map in the many-particle system as a function of the $m$-value and the colloid number density (per area), $\rho$. The $\rho$-axis is logarithmic until $\rho\sigma_0^2=1.0$ and linear above for a clearer illustration. The calculation of the correlation includes the 20 nearest neighbors of each particle. Each symbol characterizes one simulation, while dots denote liquid phases and crosses solid phases. Regions between the symbols are interpolated. The solid line confines the region of antiphase oscillations ($\nu_{\sigma\sigma,20} < 0$) and the encircled symbols denote cluster waves (for their definition, see Methods). \textbf{(c)} Amplitude of the size oscillations and \textbf{(d)} mean particle diameter in the many particle system as a function of density, $\rho$, and for different $m$, cf. legend. The amplitude is calculated according to Eq.~(\ref{eq.methods.amplitude}) (see Methods section).}
	\label{fig.results.correlations}
\end{figure*}

An isolated monostable particle cannot oscillate in size and internal proton concentration ($\sigma$-$c$-space). Such a behavior changes as soon as we add a second particle to the system. We start our analysis by looking at size-correlations between two particles at fixed distance $r$, resulting in a system with four degrees of freedom, $\sigma_{1,2}$ and $c_{1,2}$ for the particles 1 and 2. The (normalized) Pearson correlation coefficient of the two sizes is defined by
\begin{equation}
	\nu_{\sigma_1, \sigma_2} = \frac{\langle \sigma_1(t) - \langle\sigma_1\rangle\rangle \langle \sigma_2(t) - \langle\sigma_2\rangle\rangle}{\sqrt{\chi_{\sigma_1} \chi_{\sigma_2}}},
	\label{eq.methods_pcc}
\end{equation}
where $\langle \cdot\cdot \rangle$ and $\chi$ denote a time average and the variance, respectively. The former is shown in Fig.~\ref{fig.results.correlations}(a) as a heat map in the $r$-$m$-space.  For very large distances, we do not observe any correlation, because of the finite-range of the chemical communication. For closer distances, $\nu_{\sigma_1, \sigma_2}$ is nonzero and negative: a large particle produces a strong Yukawa cloud, which forces the other particle to shrink, leading to an antiphase correlation. For even shorter distances and $m \gtrsim 2 q\sigma_0 / \cc$, the Pearson correlation coefficient approaches the value -1. Here, the particles oscillate and are highly correlated in antiphase due to their mutual activation and the roughly $\pi/2$ phase-shift between size and internal proton concentration. For very short distances ($r \approx 0.2\sigma_0$) and $m \lesssim 2 q\sigma_0 / \cc$, the correlations are positive and noisy in-phase oscillations appear, which come from the interplay between the mechanical interaction (Hertzian pair potential) and the chemical feedback. However, such short interparticle distances are highly improbable in the simulations with variable positions because of the excluded-volume repulsive forces of the Hertzian pair potential. In the two-particle system, the $m$-value plays a similar role for the mechano-chemical coupling as the distance $r$. Both tune the interaction strength and therefore, for higher $m$-values, larger distances are sufficient for stable antiphase oscillations.

\subsection{Collective antiphase oscillations are tuned by colloidal density}
The two-particle dynamics has consequences on the behavior of the collective many-body dynamics. We proceed by introducing a local size correlation coefficient $\nu_{\sigma\sigma,20}$ for many diffusing and translating particles at a fixed number density (per area), $\rho$. It is similar to the Pearson correlation coefficient of Eq.~(\ref{eq.methods_pcc}), but takes into account the instantaneous 20 nearest neighbors of each particle. The averaged result as a function of number density and communication strength is shown in Fig.~\ref{fig.results.correlations}(b).

In the region of low densities, lower than the overlap concentration, $\rho\sigma_0^2 \lesssim 1$, the dilute many particle system shows a behavior in accordance with the two-particle system (cf. Fig.~\ref{fig.results.correlations}(a) versus (b)): For very low densities, $\rho\sigma_0^2 \lesssim 0.02$, the particles show little to no correlations. With increasing density, however, colloids come more often close to each other and fluctuate between isolated and nearby, communicating states. Unlike in the two-colloid system, the distance between the particles is not fixed but driven by Brownian motion and the soft Hertzian repulsion. We find especially strong anti-correlations in the sizes for close particles ($\rho\sigma_0^2 \simeq 0.3$) at moderate to large communication strength $m\gtrsim 5 q\sigma_0 / \cc$ (cf. Fig.~\ref{fig.results.correlations}(b)). We call this region \textit{antiphase oscillations}. The underlying oscillations are pseudo-regular and their amplitude varies over time. The higher the density the more frequently and more particles interact with each other at the same time, thus resulting in higher amplitudes. The average colloid size amplitude of the oscillations, plotted in Fig.~\ref{fig.results.correlations}(c), shows a peak around $\rho\sigma_0^2 \approx 0.3$. In addition, we observe an increase in the amplitude with increasing $m$ due to the mutual feedback. A significant further increase of the amplitude with $m$ is not possible, because the single particle potential $U(\sigma)$ limits the amplitude to $A < 0.5\sigma_0$.  

\subsection{Communication leads to active colloidal phases across the phase diagram}
Below the overlap concentration, $\rho\sigma_0^2 \lesssim 1$, the chemical interactions dominates. However, if we go to semi-dilute phases at higher densities, the mechanical interactions from the repulsive Hertzian pair-interaction become important. The transition from the dilute phase to the semi-dilute phase ($\rho\sigma_0^2 \gtrsim 1$) is not sharp but a crossover region, making the critical density of $\rho^\star\sigma_0^2 = 1$ a rough guide similar to existing definitions~\cite{book_Doi_Edwards}. For no coupling (vanishing $m$), the observed behavior is known from responsive colloids without chemical feedback~\cite{Baul2021}: The mean size of a particle, shown in Fig.~\ref{fig.results.correlations}(d), decreases with increasing density due to the elastic compression. However, an increase in $m$ intensifies this effect because the chemical signaling of the neighboring particles leads in average to a shrinking, too. The peak in amplitude (cf. Fig.~\ref{fig.results.correlations}(c)) must therefore be assigned to the chemical coupling in the system. Again, we see how the communication tunes nonequilibrium structuring of the colloidal dispersion. 

For very high densities ($\rho\sigma_0^2 \gtrsim 2.4$) and low chemical coupling the system transitions into a solid hexagonal lattice phase (cf. crosses in Fig.~\ref{fig.results.correlations}(b)). Systems in the solid phase are identified by an extended hexagonal order parameter (see Methods section). Nonequilibrium effects can also be clearly observed here: The critical (`phase transition') density for the solid phase increases with $m$, since the particle size is overall decreased by their chemical response (cf. Fig.~\ref{fig.results.correlations}(d)).
{In contrast, compared to equilibrium solid phase ($m=0$), colloids in this {\it active} solid phase are slightly larger (see Fig.~\ref{fig.results.correlations}(d) and Supplementary Fig.~S1(f)). The explanation lies in the interplay between the active fluctuations induced by the chemical coupling and the Hertzian pair interaction in the dense solid. In addition to the size effect, we observe a small reduction of long-range structure for increasing $m$ (cf. Supplementary Fig.~S2(f)).}

\begin{figure}[h!]
	\centering
	\includegraphics[width=8.6cm]{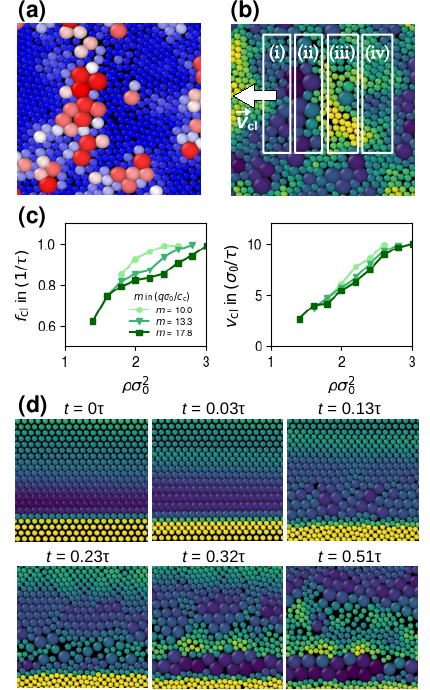}
	\caption{{\bf In-phase cluster oscillations and elastic waves.} \textbf{(a)} Simulation snapshot of a cluster oscillation. The colors indicate the size of a particle. \textbf{(b)} Same snapshot as in (a), but the colors now decode the measured chemical signal of each particle (own concentration + signal of all other particles), where bright colors indicate a high signal (proton concentration). The regions (i)-(iv) show the different states a particle passes through during one oscillation cycle (cf. Fig.~\ref{fig.methods}(a)). The arrow indicates the propagation direction of the wave with velocity $\vec v$. \textbf{(c)} Frequency (left), and velocity (right) of the cluster waves as a function density, $\rho$, for different $m$-values. \textbf{(d)} Time series of simulation snapshots of a simulation with an inhomogeneous initial concentration distribution and in the absence of noise. The time series shows the forming of a cluster wave in vertical direction. As in (b) the colors decode the currently measured signal of a particle. Related movies can be found in the Supplementary Movies S5-S7.}
	\label{fig.results.cluster}
\end{figure}

Hence, we emphasize here that irrespective of the presence of oscillations, the communication $m$ significantly modifies the structure and dynamics in the {\it active} colloidal liquid at all densities (in all phases) compared to the equilibrium colloidal liquid with $m=0$. This concerns, for example, the polydispersity in the size distribution (Supplementary Fig.~S1) or the radial distribution functions, $g(r)$ (Supplementary Fig.~S2) across all densities. 

\subsection{Cluster oscillations and elasto-chemical cluster waves}
For high communication strength $m$ and densities above $\rho\sigma_0^2 \approx 1.0$ the interplay between chemical and mechanical interactions as well as elastic packing amplifies and new qualitative physics is observed in the liquid phase, which we call \textit{cluster waves} (see encircled symbols in Fig.~\ref{fig.results.correlations}(b)). A simulation snapshot of such a cluster oscillation wave is shown in Figs.~\ref{fig.results.cluster}(a) and (b) (see also Supplementary Movies S5, S6). In the range of one wavelength (large particles - small particles - large particles) we find particles in all four states (i)-(iv) of the overdamped single-particle oscillations from Fig.~\ref{fig.methods}(a): (i) Recently swollen particles which measure therefore a low signal. (ii) Long time swollen particles which will shrink soon due to the high signal. (iii) Recently collapsed particles which have therefore still a high concentration of the measured chemical signal. (iv) Collapsed particles with low signal which will grow as soon as they get space for it. They will acquire the space from the now state-(ii) particles which will flow into the free volume of the collapsing state-(i) particles. In total, we obtain a longitudinal wave similar to a sound wave. Note that the particles' translational movement is essential for the appearance of cluster waves. Nevertheless, after one oscillation period the particles change their position only marginally; there is no net flux of particles. Using this, we define a system to be in the cluster wave regime if the short-term diffusion is partially superdiffusive (for details see Supplementary Fig.~S3). With this definition we find cluster waves for $\rho\sigma_0 \geq 1.4$. However, the transition from anti-phase oscillations to cluster waves is smooth, and the cluster size and stability increase with density (and $m$). The oscillating properties in such a cluster wave are the chemical signal (proton) field, particle size, and local number density.

In the following, we analyze deeper the properties of these cluster oscillations.  The cluster frequency, $f_\mathrm{cl}$, in Fig.~\ref{fig.results.cluster}(c) shows the density range in which waves are possible. In accordance with Fig.~\ref{fig.results.correlations}(b), higher $m$ values enable cluster waves at higher densities. This follows again from the increasing signal response at higher $m$, which in turn leads to smaller particles and therefore a higher critical freezing density. Furthermore, we find a slightly increasing frequency with increasing density and decreasing $m$-parameter. 

To obtain the cluster wave phase velocity, we combine the frequency $f_\mathrm{cl}$ and wavelength $\lambda_\mathrm{cl}$ via the dispersion relation for the phase velocity, $v_\mathrm{cl} = \lambda_\mathrm{cl} f_\mathrm{cl}$; the result is shown in Fig.~\ref{fig.results.cluster}(c). We observe a linear increase of $v_\mathrm{cl}$ with $\rho$. We suppose that the increased interparticle elastic interactions press the particles faster into free volumes and thereby speed up the wave phase velocity. The observed cluster waves are typically undirected and many small cluster waves overlap and run into each other. However, by switching the noises off and starting in an initial configuration with concentration gradient, we can produce temporally limited directed cluster waves. Figure~\ref{fig.results.cluster}(d) shows simulation snapshots from a movie of such a directed wave (see Supplementary Movie S7).

\section{Discussion}
We demonstrate how chemical communication between size-responsive (hydrogel) particles affects the structure and phase behavior of active colloidal suspensions. In our model, the colloids respond to each other's chemical clouds, which was inspired by the glucose-proton chemo-mechanical feedback cycle~\cite{Wessling2021} and the chemical signaling mechanism of cells and bacteria~\cite{QS, Bauerle2018}. We demonstrated that the communication can lead to numerous very distinct dynamical states with novel active features tuned by the chemical responsiveness and density. Even though isolated colloids were not able to oscillate, we found collective antiphase oscillations (high chemical response, low density) and cluster waves (high chemical response, high density). Thereby, one crucial point lies in the interplay between chemical response and mechanical response. Another point is the importance of threshold densities which is also demonstrated in the glycolytic oscillations of yeast cells~\cite{DeMonte2007}, and chemo-mechanical oscillations in active hydrogels with feedback to the Belousov-Zhabotinsky reaction~\cite{Blanc2024}.

While cell-cell communication and rich dynamics is inherent to biological matter, artificial systems based on enzymatically active hydrogels have been synthesized and demonstrated to show oscillatory behavior~\cite{Wessling2021}, which can, in principle, with further bulk synthesis and fine-tuning verify our results. We believe that the explicit communication between two of such colloids might be testable in optical tweezers experiments of two mono- or bistable hydrogel colloids~\cite{Rica}. Furthermore, artificial quorum sensing was demonstrated also through porous artificial cell-mimics containing a nucleus-like DNA-hydrogel compartment that is able to express and display proteins, and communicate with neighbors  through diffusive protein signals~\cite{Niederholtmeyer2018}.  In another approach, giant vesicle-based synthetic cells which, via gene expression, synthesized in their aqueous lumen an enzyme that in turn produces a chemical signal then perceived by a bacterium~\cite{Rampioni2018}.  Systematically extending such experimental works from the perspective of our contribution could be used directly to test our prediction. We believe that especially the cluster waves are important for information flow in synthetic nonequilibrium materials to create adaptive materials with life-like properties~\cite{C6CS00738D,Niederholtmeyer2018,Nguyen2018}. Here, further research is necessary to generate longer-living and more directed waves. This could possibly be realized with additional external and steering chemical fields or confining boundary conditions.

Theoretically, our work should inspire future developments in the nonequilibrium physics of active chemo-mechanically coupled liquids~\cite{Frey}. Desirably and appropriate would be, in particular, hydrodynamic field theories, as put forward already for motile-active colloids (swimmers)~\cite{Witti,stark,Dinelli2023,Bricard2013}. The results here should also be interesting for questions regarding fluctuation-dissipation and chemical work in the emerging field of stochastic thermodynamics~\cite{Seifert_2012,PhysRevX.14.011012} or in the emerging physics of chemically-driven active emulsions~\cite{Weber_2019}. Furthermore, the communication fields (in Eq.~(1)) are intrinsically non-Hamiltonian and our system therefore has so-called non-reciprocal interactions (violating Newtons third law). Non-reciprocity leads to a rich set of emerging nonequilibrium behaviors for which the development of theoretical frameworks are challenging~\cite{Kolmakov2010,Loos2020,Dinelli2023,Kreienkamp_2022}. Finally, we wonder whether it is in future possible to coarse-grain some features of our more microscopic model to locally coupled Kuramoto oscillators~\cite{Wuster2020}, recently used to model pulsating active matter~\cite{Tjhung2017,Zhang2023}.

\section{Methods}
\subsection{Model}
	
We consider a two-dimensional (2D) system of $N$ particles where each particle $i$ has four degrees of freedom (DoFs): position $\boldsymbol{x}_i$, diameter $\sigma_i$, and inside concentration $c_i$. All particles are assumed to be spherical (3D) while fixed to the $x$-$y$-plane (cf. Fig.~\ref{fig.methods}(c)). The translational and size dynamics are governed by overdamped (BD) equations of motion, while we use a chemical rate equation for the concentration DoF (cf. Eqs.~(\ref{eq.model_eom_x})-(\ref{eq.model_eom_c})).

The BD equations (\ref{eq.model_eom_x}) and (\ref{eq.model_eom_s}) employ a friction term with friction coefficients $\xi_x$ and $\xi_\sigma$, respectively. For the translational friction coefficient we use Stokes' law which leads to an instantaneous friction coefficient of $\xi_x(\sigma) = \xi_x^0 \sigma/\sigma_0$ with $\xi_x^0=1 \kBT\tau/\sigma_0^2$ being the reference friction coefficient of a particle with size $\sigma_0$ and defining the unit time $\tau$. The size friction coefficient is fixed to $\xi_\sigma=10\xi_x^0$. For simplicity, we neglect hydrodynamic interactions. The translational motion, Eq.~(\ref{eq.model_eom_x}), additionally comprises a particle-particle interaction term and a noise term. For the interaction we use the purely repulsive (elastic) Hertzian pair-potential%
\begin{equation}
	\Phi_{ij} = \epsilon \left(1 - \frac{r_{ij}}{\sigma_{ij}}\right)^{5/2} \Theta\left(1 - \frac{r_{ij}}{\sigma_{ij}}\right)
\end{equation}
where $\Theta(\cdot\cdot)$ denotes the Heaviside step function, $\epsilon$ is the potential strength, $r_{ij}$ is the center-to-center distance of the two particles $i$ and $j$, and $\sigma_{ij} = (\sigma_i + \sigma_j)/2$ is the mean diameter of the two particles. Experiments show that $\epsilon = 500 k_\mathrm{B} T$ is a typical value for thermosensitive colloids~\cite{Paloli2013}. The repulsive Hertzian potential is meant to avoid very short distances between particles. To take the interactions with the surrounding solvent into account a Gaussian noise term $\eta_{ix}$ is used with $\langle \eta_{ix}(t)\rangle = 0$ and $\langle \eta_{ix}(t) \eta_{jx}(t')\rangle = \delta_{ij} \delta(t-t') 2 k_\mathrm{B}T \xi_x / \Delta t$. Here, $k_\mathrm{B}$ is the Boltzmann constant, $T$ the temperature of the solvent, $\Delta t$ the simulation timestep, and $\delta_{ij}$ and $\delta(t-t')$ denote the Kronecker delta and Dirac $\delta$-function, respectively.

Eq.~(\ref{eq.model_eom_s}) describes changes of the particle's size $\sigma_i$ and couples all differential equations. The first and the last term are analogies to Eq.~(\ref{eq.model_eom_x}). The last term is a noise and adds fluctuations to the size. The first term is a derivative of the Hertzian potential with respect to $\sigma_i$ and leads to shrinking of overlapping particles. The second term is a force from the single-particle potential of the FENE (finite extensible nonlinear elastic~\cite{fene}) form
\begin{eqnarray}
	U(\sigma) &=& -\frac{q \sigma_0^2}{8 \beta} \ln\left[1 - 4\left(\frac{\sigma - \sigma_0}{\sigma_0}\right)^2\right].
	\label{eq.single-particle-pot}
\end{eqnarray}
Therefore, we have a preferred/equilibrium size $\sigma_0$ and in first order a Hooke's law for the force towards the equilibrium size $\sigma_0$ with (spring) constant $q = 100\kBT/\sigma_0^2$. Only for very small and very large sizes $U(\sigma)$ deviates from the harmonic potential to limit the colloids diameter to $0.5\sigma_0 < \sigma < 1.5\sigma_0$. Thus, the colloids can neither vanish nor increase tremendously.

Finally, Eq.~(\ref{eq.model_eom_c}) is the rate equation of the particle's concentration $c_i$. The first term describes the production of protons from a constant fuel (e.g., glucose) which is proportional to a rate $k_0=2.749/\tau$ (including the constant glucose concentration $g$). In addition, it depends exponentially on the particle's size with the sieving parameter $B=1\sigma_0^3$. The larger the particle the higher is the concentration production. This is motivated by active hydrogels, where the hydrogel's membrane can be permeable for glucose if it is swollen and is impenetrable in the collapsed state~\cite{Wessling2021}. In this analogy is $c_i$ the proton concentration inside the particle, emerging as product from a chemical reaction of glucose inside the hydrogel. The second term in Eq.~(\ref{eq.model_eom_c}) is a simple loss term which is proportional to the \textit{ad hoc} concentration and a loss rate $k_l=1/\tau$. A similar kinetic equation was used in~\cite{Milster2023} to model the nonlinear dynamic of feedback-controlled microreactors. The proton production rate $k_0=2.749/\tau$ is chosen in such a way that the size distribution of an isolated particle does not change with $m$.

The communication between the particles is described by the third term of Eq.~(\ref{eq.model_eom_s}). We assume that the loss of the proton concentration inside a particle is due to diffusion. We further assume that this proton diffusion is free through space, because of their small size, and therefore create a proton cloud around each particle. Furthermore, we approximate that proton diffusion happens much faster than all other timescales and thus only depends on the (glucose) concentration and size at the current time, $c_i$ and $\sigma_i$. The ansatz for an instantaneous chemical field was chosen in~\cite{Fadda2023}. We presume that the proton cloud is well characterized by a Yukawa function, which is the stationary solution of the diffusion equation for a fixed and constant point source with loss term~\cite{LiebchenBook} and hence a well-known ansatz~\cite{Zampetaki2021}. Consequently, the proton field a particle $i$ feels from particle $j$ in a center-center distance $r_{ij}$ is approximated by
\begin{equation}
	Y_{ji}(r_{ij}; c_j, \sigma_j) = c_j \frac{\sigma_j e^{\kappa \sigma_j/2}}{2} \frac{1}{r_{ij}} e^{-\kappa r_{ij}}
	\label{eq.yukawa}
\end{equation}
with an inverse signal range $\kappa$. We fix the communication length to the particle scale $\kappa^{-1}= \sigma_0$. The signal a particle sends out is proportional to its own (glucose) concentration $c_j$, since we assume a chemical reaction from glucose to protons. In addition, the signal depends on the particle diameter such that the signal at the particle's surface is equal to the inside concentration, $Y_{ji}(r_{ij}=\sigma_j/2; c_j, \sigma_j) = c_j$.

In total, every particle senses the sum of all other proton fields at its center, plus the own inside concentration (self-sensing). Finally, we assume that there is a size-independent `ideal' proton concentration $\cc$ for each particle to sense. If the sensed proton concentration, $ c_i + \sum_{j \neq i}Y_{ji}$, is higher, the particle will shrink and \textit{vice versa}. This again is motivated by the behavior of active hydrogels at different pH values~\cite{Wessling2021}. The response strength is characterized by $m$.

\subsection{Simulation}
The simulation consists of 1600 particles in a rectangular box with periodic boundary conditions. A simulation snapshot is shown in Fig.~\ref{fig.methods}(c). The equations of motion of Eqs.~(\ref{eq.model_eom_x})-(\ref{eq.model_eom_c}) are integrated with the fourth order Runge-Kutta scheme  and timestep of $\Delta t = 10^{-4}\tau$. The simulation is started from an instable configuration (initial concentration gradient, see Fig.~\ref{fig.results.cluster}(d)) and equilibrates for $5\tau$, before data is collected for a duration of $40\tau$.

The FENE single-particle potential $U(\sigma)$ diverges at $\sigma_{1,2}^\mathrm{lim} = 0.5\sigma_0,~1.5\sigma_0$. If the diameter nevertheless leaves this region, it is reset to $\sigma = 0.501\sigma_0$ or $\sigma = 1.499\sigma_0$, respectively. The same is done if an inside concentration becomes negative ($c^\mathrm{new} = 0.01\cc$).

A table with all simulation parameters is shown in the Supplementary Table~S1.

\subsection{Order parameters}
{\it Solid phase hexagonal order --} To quantify the state diagram region which is in the solid phase (see Fig.~\ref{fig.results.correlations}(b)), we use an extension of the conventional hexagonal order parameter~\cite{Gasser2010}, which we define as
\begin{equation}
	\Psi_{18} = \left\langle \left| \frac{1}{18} \sum_{k=19}^{36} e^{18i\Theta_{jk}} \right| \right\rangle,
\end{equation}
where the sum runs over the nearest neighbors 19 to 36 of a particle $j$. Here, $\Theta_{jk}$ is the angle formed by the $x$-axis and the connecting line between the particles $j$ and $k$. The parameter $\Psi_{18}$ is 1 in case of a perfect hexagonal lattice and approaches $0.210$ for entirely uncorrelated positions. A plot of $\Psi_{18}$ as a function of density and communication strength is shown in Supplementary Fig.~S4. We define a system to be in the solid phase if $\Psi_{18} > 0.5$.

\vspace{0.5cm}
{\it Amplitude --} Other than in common oscillations, it is here not trivially possible to read out the amplitude from the trajectory of a particle $\sigma_i(t)$, because the oscillations occur pseudo-randomly in time, with different intensities, and are affected by noise. To overcome these problems, we decided to split the trajectory into non-overlapping time intervals of length $2\tau$, and determine the minimum and maximum value of $\sigma(t)$ in this interval, $\sigma_\mathrm{min}$ and $\sigma_\mathrm{max}$. The amplitude is defined as the ensemble and time average over the non-overlapping time intervals of
\begin{equation}
	A = \left\langle (\sigma_\mathrm{max} - \sigma_\mathrm{min})/2 \right\rangle.
	\label{eq.methods.amplitude}
\end{equation}
Due to the fluctuations in size from the noise term, it always holds $A > 0$; even for an isolated particle where no oscillations are possible.

\vspace{0.5cm}
{\it Frequency, wavelength, and velocity.--} The goal is to obtain the frequency and the wavelength of the cluster oscillations. Therefore, we use the total signaling field, which we calculate for one point in space-time by summing up the Yukawa clouds of all particles $Y_\mathrm{tot}(\mathbf{r}, t) = \sum_{i} Y_i(|\mathbf{r}-\mathbf{r}_i|, t)$. To avoid a divergence in a particle's center, we fix the Yukawa cloud inside the particle ($|\mathbf{r}-\mathbf{r}_i| < \sigma_i/2$) to $Y_i(|\mathbf{r}-\mathbf{r}_i|, t) = c_i$. Note that this is exactly the value at the surface which guarantees continuity of the Yukawa cloud. We calculate the signaling field every $0.025\tau$ on a 20x20 rectangular grid in space. We normalize the total Yukawa cloud to a unit-less quantity with zero mean $\tilde{Y}_\mathrm{tot} = (Y_\mathrm{tot} - \langle Y_\mathrm{tot}\rangle)/\cc$ and calculate its auto-correlation functions in space and time
\begin{eqnarray}
	C_r(\mathbf{r}, t=0) &=& \left\langle \tilde{Y}_\mathrm{tot}(\mathbf{r}, 0) \tilde{Y}_\mathrm{tot}(0, 0) \right\rangle, \\
	C_t(\mathbf{r}=0, t) &=& \left\langle \tilde{Y}_\mathrm{tot}(0, t) \tilde{Y}_\mathrm{tot}(0, 0) \right\rangle,
	\label{eq.methods.vanHove}
\end{eqnarray}
where $\langle \cdot \cdot \rangle$ denotes a time and ensemble average ($\mathbf{r}=0$ can be every point in space). This is justified because we assume isotropy in space. With the same reasoning we reduce the spatial dependency to a distance $C_r(r=|\mathbf{r}|, t)$. Subsequent, we calculate the Fourier transforms of the auto-correlation functions, $\mathcal{F}[C_r(r)]$ and $\mathcal{F}[C_t(t)]$, which reveals the occurring wave numbers and frequencies. We determine the maxima in $\mathcal{F}[C_r(r)](k)$ and $\mathcal{F}[C_t(t)](\omega)$, and therewith define the cluster wavelength $\lambda_\mathrm{cl} = 2\pi/k_\mathrm{max}$, cluster frequency $f_\mathrm{cl} = 2\pi/\omega_\mathrm{max}$, and cluster velocity $v_\mathrm{cl} = \lambda_\mathrm{cl} f_\mathrm{cl}$. The discrete Fourier transforms are interpolated quadratically between the three closest points to the maximum to obtain a more precise maximum position. By estimating the uncertainty as half the distance between two points in Fourier space, we get relative uncertainties of up to $10\%$ for the frequency and up to $15\%$ for the wave velocity.


\section{Acknowledgments}
\begin{acknowledgments}
	We thank Sebastian Milster, Sebastien Groh, and Wolfgang Boemke for useful discussions. The authors also acknowledge the support by the state of Baden-W\"urttemberg through bwHPC and the DFG through Grant No. INST 39/963-1 FUGG (bwForCluster NEMO) under Germany's Excellence Strategy-EXC-2193/1-390951807 (``LivMatS").
\end{acknowledgments}




\providecommand{\noopsort}[1]{}\providecommand{\singleletter}[1]{#1}%


%

%

%

%
%
	
\end{document}